\documentclass[twocolumn,showpacs,prc,superscriptaddress,amsmath,amssymb]{revtex4-1}
\usepackage{multirow}
\usepackage{graphicx}% Include figure files
\usepackage{dcolumn}% Align table columns on decimal point
\usepackage{bm}% bold math
\usepackage{soul}
\usepackage{color}
\begin{document}

\preprint{APS/123-QED}

\title{First observation of $^{13}$Li ground state}

\author{Z. Kohley}
 \email{kohley@nscl.msu.edu}
\affiliation{National Superconducting Cyclotron Laboratory, Michigan State University, East Lansing, Michigan 48824, USA}
\author{E. Lunderberg}
\affiliation{Department of Physics, Hope College, Holland, Michigan 49422, USA}
\author{P.\ A.\ DeYoung}
\affiliation{Department of Physics, Hope College, Holland, Michigan 49422, USA}
\author{A. Volya}
\affiliation{Department of Physics, Florida State University, Tallahassee, Florida 32306, USA}
\author{T. Baumann}
\affiliation{National Superconducting Cyclotron Laboratory, Michigan State University, East Lansing, Michigan 48824, USA}
\author{D. Bazin}
\affiliation{National Superconducting Cyclotron Laboratory, Michigan State University, East Lansing, Michigan 48824, USA}
\author{G. Christian}
\altaffiliation[Present address: ]{TRIUMF, 4004 Wesbrook Mall, Vancouver, British Columbia V6T 2A3, Canada}
\affiliation{National Superconducting Cyclotron Laboratory, Michigan State University, East Lansing, Michigan 48824, USA}
\affiliation{Department of Physics $\&$ Astronomy, Michigan State University, East Lansing, Michigan 48824, USA}
\author{N. L. Cooper}
\affiliation{Department of Physics and Astronomy, Indiana University at South Bend, South Bend, Indiana 46634, USA}
\author{N. Frank}
\affiliation{Department of Physics and Astronomy, Augustana College, Rock Island, Illinois 61201, USA}
\author{A. Gade}
\affiliation{National Superconducting Cyclotron Laboratory, Michigan State University, East Lansing, Michigan 48824, USA}
\affiliation{Department of Physics \& Astronomy, Michigan State University, East Lansing, Michigan 48824, USA}
\author{C. Hall}
\affiliation{Department of Physics, Hope College, Holland, Michigan 49422-9000, USA}
\author{J. Hinnefeld}
\affiliation{Department of Physics and Astronomy, Indiana University at South Bend, South Bend, Indiana 46634, USA}
\author{B. Luther}
\affiliation{Department of Physics, Concordia College, Moorhead, Minnesota 56562, USA}
\author{S. Mosby}
\altaffiliation[Present address: ]{Los Alamos National Laboratory, Los Alamos, New Mexico 87545, USA}
\affiliation{National Superconducting Cyclotron Laboratory, Michigan State University, East Lansing, Michigan 48824, USA}
\affiliation{Department of Physics $\&$ Astronomy, Michigan State University, East Lansing, Michigan 48824, USA}
\author{W. A. Peters}
\altaffiliation[Present Address: ]{Oak Ridge Associated Universities, Oak Ridge, TN 37830, USA}
\affiliation{National Superconducting Cyclotron Laboratory, Michigan State University, East Lansing, Michigan 48824, USA}
\affiliation{Department of Physics $\&$ Astronomy, Michigan State University, East Lansing, Michigan 48824, USA}
\author{J. K. Smith}
\affiliation{National Superconducting Cyclotron Laboratory, Michigan State University, East Lansing, Michigan 48824, USA}
\affiliation{Department of Physics $\&$ Astronomy, Michigan State University, East Lansing, Michigan 48824, USA}
\author{J. Snyder}
\affiliation{National Superconducting Cyclotron Laboratory, Michigan State University, East Lansing, Michigan 48824, USA}
\affiliation{Department of Physics $\&$ Astronomy, Michigan State University, East Lansing, Michigan 48824, USA}
\author{A. Spyrou}
\affiliation{National Superconducting Cyclotron Laboratory, Michigan State University, East Lansing, Michigan 48824, USA}
\affiliation{Department of Physics $\&$ Astronomy, Michigan State University, East Lansing, Michigan 48824, USA}
\author{M. Thoennessen}
\affiliation{National Superconducting Cyclotron Laboratory, Michigan State University, East Lansing, Michigan 48824, USA}
\affiliation{Department of Physics $\&$ Astronomy, Michigan State University, East Lansing, Michigan 48824, USA}
\date{\today}% It is always \today, today,

\begin{abstract}
The ground state of neutron-rich unbound $^{13}$Li was observed for the first time in a one-proton removal reaction from $^{14}$Be at a beam energy of 53.6~MeV/u. The $^{13}$Li ground state was reconstructed from $^{11}$Li and two neutrons giving a resonance energy of 120$^{+60}_{-80}$~keV. All events involving single and double neutron interactions in the Modular Neutron Array (MoNA) were analyzed, simulated, and fitted self-consistently.  The three-body ($^{11}$Li+$n$+$n$) correlations within Jacobi coordinates showed strong dineutron characteristics. The decay energy spectrum of the intermediate $^{12}$Li system ($^{11}$Li+$n$) was described with an $s$-wave scattering length of greater than $-4$~fm, which is a smaller absolute value than reported in a previous measurement.

\end{abstract}

                             % Classification Scheme.
%\keywords{Suggested keywords}%Use showkeys class option if keyword

\maketitle
The increasing availability of rare isotope beams has made it possible to extend nuclear structure measurements to nuclei far away from stability.   In light neutron-rich nuclei these studies have been extended beyond the neutron dripline~\cite{Bau12}. The existence of two-neutron halos, first observed in $^{11}$Li~\cite{Tan85,Han87}, sparked the interest in unbound two-neutron systems where measurements of the decay products could yield information about neutron-neutron correlations inside the nucleus. Recent examples include $^{10}$He~\cite{Joh10,Joh10a}, $^{13}$Li~\cite{Joh10a,Aks08}, $^{16}$Be~\cite{Spy12}, and $^{26}$O~\cite{Lun12}. In $^{16}$Be the first evidence for ground state dineutron decay was observed because sequential decay via $^{15}$Be was energetically not possible~\cite{Spy12}.

%Recently these experiments have been pushed beyond oxygen with the first mass measurement of unbound $^{28}$F~\cite{Chr12}.

In this article, we report on a new measurement of the $^{13}$Li decay products. Very few calculations exist that predict the energy levels of $^{13}$Li. Over 25 years ago, a shell model calculation predicted the ground state of $^{13}$Li to have a spin and parity of $\frac{3}{2}^-$ located about 3 MeV above the $^{11}$Li+2$n$ threshold~\cite{Pop85}. In 2008 Aksyutina {\it et al.}~\cite{Aks08} reported a resonance at a decay energy of 1.47(31)~MeV in a proton removal reaction from $^{14}$Be at 304~MeV/u with a liquid hydrogen target. In the same reaction the unbound intermediate nucleus $^{12}$Li was observed to decay via the emission of an $s$-wave neutron with a scattering length of $-$13.7(1.6)~fm. We repeated this proton removal reaction at a $^{14}$Be beam energy of 53.6~MeV/u on a beryllium target. In addition to measuring a new low-lying resonance in $^{13}$Li at 120$^{+60}_{-80}$~keV, which is presumably the ground state, we extract a limit on the scattering length of $a_{s} > -4$ fm for $^{12}$Li.

The experiment was performed at the Coupled Cyclotron Facility of the National Superconducting Cyclotron Laboratory (NSCL) at Michigan State University. A 120~MeV/u $^{18}$O primary beam bombarded a 4113~mg/cm$^{2}$ Be production target to produce the 53.6~MeV/u $^{14}$Be secondary beam.  The A1900 fragment separator, with a 1050~mg/cm$^{2}$ Al achromatic wedge degrader, was used to separate the $^{14}$Be secondary beam from other reaction products and the primary beam. The rate of the $^{14}$Be secondary beam was about 500 pps at the reaction target and had a momentum spread of 2.5\%.  Contaminant nuclei in the secondary beam were excluded from analysis based on event-by-event measurement of the time-of-flight from plastic scintillator timing detectors located at the A1900 focal plane and just before the reaction target. The $^{14}$Be beam impinged on a 477~mg/cm$^2$ Be reaction target and produced $^{13}$Li through one-proton removal. The two-neutron unbound $^{13}$Li decayed immediately into $^{11}$Li and two neutrons.

The $^{11}$Li fragments were deflected by the Sweeper superconducting dipole magnet~\cite{Bir05} that was set to a magnetic rigidity of 3.735~Tm. The detection setup was identical to the experiment by Hall {\it et al.}~\cite{Hal10}, which measured excited states in $^{12}$Li following the two-proton removal from $^{14}$B. The deflected nuclei passed through two x-y position sensitive cathode-readout drift chambers (CRDCs), separated by 1.816~m, a 5~mm thick plastic scintillator, and were stopped in a 150~mm thick plastic scintillator. An inverse ion-optical matrix for ray tracing was created with the program \textsc{cosy infinity}~\cite{Mak05} using the measured magnetic field map of the Sweeper to provide tracking of the $^{11}$Li trajectories back to the target position.

The neutrons produced from the $^{13}$Li breakup were detected using the Modular Neutron Array (MoNA)~\cite{Lut03,Bau05}. The individual plastic scintillator bars were arranged in nine walls, each 16 bars high, resulting in an active area 2~m wide by 1.6~m tall. The distance from the target to the center of the first layer was 8.44~m.

%The three-body decay energy of $^{13}$Li can be calculated with $E_{\mbox{\scriptsize Decay}}=m_{^{13}{Li}}-m_{^{11}{Li}}-2m_{n},$
The three-body decay energy of $^{13}$Li can be calculated with $E_{\mathrm{Decay}} = M_{^{13}\mathrm{Li}} - M_{^{11}\mathrm{Li}} - 2M_{\mathrm{n}}$, where $M_{^{13}\mathrm{Li}}$ ($M_{^{11}\mathrm{Li}}$) is the mass of $^{13}$Li ($^{11}$Li) and $M_{\mathrm{n}}$ is the neutron mass.  The invariant mass, $M_{^{13}\mathrm{Li}}$, was calculated from the experimentally measured four-momenta of the $^{11}$Li and two neutrons.  It should be noted that all experimental decay energy spectra are presented without unfolding of the resolution or acceptance.

%$E_{\mbox{\scriptsize Decay}}=m_{0}-m_{f}-2m_{n}$, where $m_{0}$ ($m_{f}$) is the mass of $^{13}$Li ($^{11}$Li).  The invariant mass of $^{13}$Li can be determined from the reconstructed four-momentum of the $^{11}$Li fragment and two neutrons:
%\begin{eqnarray}
%m^2_{0} &= &m^2_{f}+2m^2_{n} + \nonumber \\
%& & \mbox{} 2(E_{f}E_{n_1}+E_{f}E_{
%n_2}+E_{n_1}E_{n_2}) - \nonumber \\
%& & \mbox{} 2(p_{f}p_{
%n_1}\cos\theta_{f, n_1}+p_{f}p_{
%n_2}\cos\theta_{f, n_2} \nonumber \\
%& & \mbox{} +p_{n_1}p_{
%n_2}\cos\theta_{n_1, n_2})
%\label{eq:invariant}
%\end{eqnarray}
%where $m$ is the rest mass of the respective particle, $E$ is the total energy, $p$ is the magnitude of the momentum, and $\theta$ is the angle between the two indicated (subscripted) particles.  It should be noted that all experimental decay energy spectra are presented without unfolding of the resolution or acceptance.

\begin{figure}
	\centering
	\includegraphics[width=0.35\textwidth]{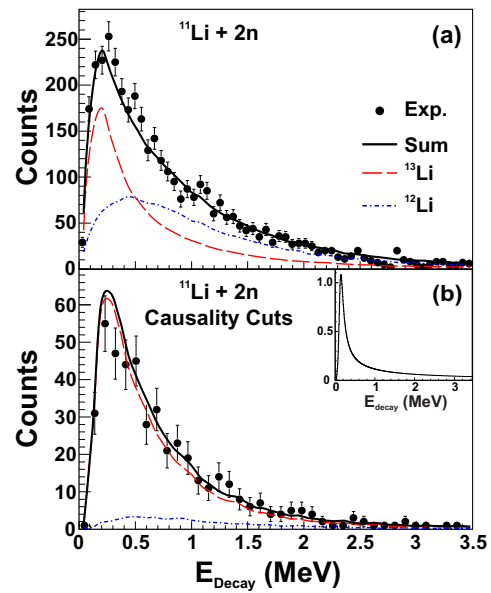}
	\caption{(color online) Measured three-body decay energy spectrum of $^{11}$Li+2$n$ without (a) and with (b) causality cuts as described in the text.  The experimental data is fit with two components: (1) the dineutron decay of $^{13}$Li$\rightarrow^{11}$Li+2$n$ (red long-dashed line) and (2) the $^{12}$Li$\rightarrow^{11}$Li+$n$ decay (blue dot-dashed line)  The input $^{13}$Li resonance distribution is shown in the insert of panel (b).
\label{f:13Li}}
\end{figure}

The reconstruction of the $^{13}$Li decay energy relies on the correct identification of two neutrons interacting in MoNA. It is critical to distinguish real two-neutron events from events where a single neutron re-scatters and interacts twice in MoNA. It has been shown that causality cuts applied to the distance and the velocity between two successive interactions can effectively enhance the selection of real two-neutron events~\cite{Spy12,Lun12,Nak06,Hof11}. Fig.~\ref{f:13Li} shows the reconstructed three-body decay energy spectrum of $^{13}$Li ($^{11}$Li+2$n$) without (a) and with (b) the causality cuts applied.   The cuts require that two successive interactions be separated by more than 50~cm and that the time between the first and second interaction be shorter than the time it would take a neutron at beam velocity to travel between the respective interaction points (see Ref.~\cite{Koh12} and references therein for additional information on 2$n$ cuts).  The data exhibit a resonance-like structure below 500~keV, which was not observed in the previous measurement of $^{13}$Li~\cite{Aks08}. However, the earlier experiment had zero efficiency for the detection of two-neutron events below 200~keV~\cite{Aks09}, which was also shown to have significant effects on the reconstructed shape of the decay energy spectrum of the excited $^{11}$Li~\cite{Nak06}.
%which also had significant effects on the reconstructed shape of the decay energy spectrum of the excited $^{11}$Li~\cite{Nak06}.

\begin{figure*}
	\centering
	\includegraphics[width=.8\textwidth]{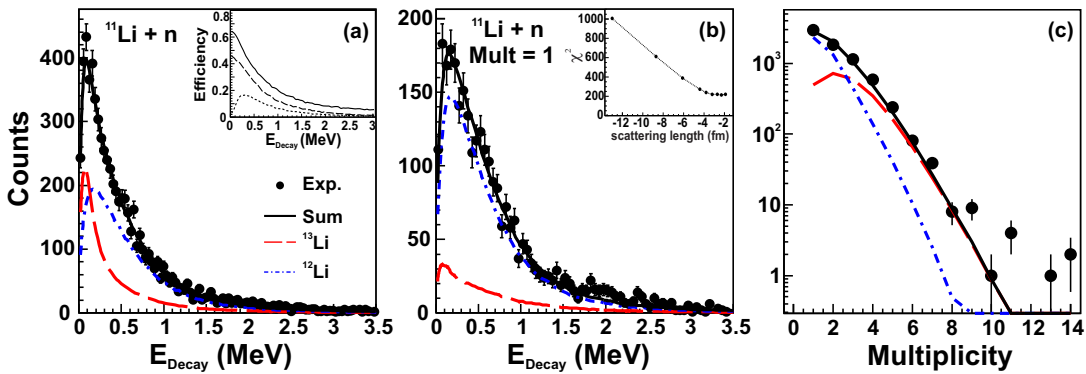}
	\caption{(color online) Measured decay energy spectrum of $^{11}$Li+$n$ including all multiplicities (a) and with a multiplicity~$=$~1 gate (b). Panel (c) displays the overall multiplicity distribution for all events in coincidence with a $^{11}$Li fragment.  The insert in panel~(a) shows the efficiency for detecting one neutron from a 1$n$ decay (solid line) and two neutrons from a 2$n$ decay with (dashed line) and without (long-dashed line) the causality cuts.  The 2$n$ decay was simulated with each neutron being emitted with 50$\%$ of the total decay energy.  The $\chi^{2}$ as a function of the $^{12}$Li $s$-wave scattering length is shown in the insert of panel~(b).
\label{f:12Li}}
\end{figure*}

%In order to extract the resonance energy of the observed peak, detailed Monte Carlo simulations were performed~\cite{Koh12}.

In order to analyze the observed peak for the presence of a resonance, detailed Monte Carlo simulations were performed~\cite{Koh12}. The simulations included the incoming beam characteristics, the reaction mechanism to populate specific states, and their subsequent decay as well as the detector resolutions and efficiencies.  The one-proton and $pn$-removal reactions were simulated using the Glauber reaction model.  As mentioned earlier the detailed response of MoNA to the interactions of multiple neutrons is especially important and was simulated with \texttt{\sc Geant4}~\cite{Ago03,All06} using the custom neutron interaction model \textsc{menate\_r}~\cite{Roe08}. In contrast to the standard intranuclear cascade models in \texttt{\sc Geant4}, which uses only total inelastic reaction cross sections for neutrons above 20~MeV, \textsc{menate\_r} uses cross sections for the different reaction channels, including ($n,np$), ($n,p$), ($n,n\gamma$), and ($n,\alpha$). It has recently been shown that the use of \textsc{menate\_r} is important for correctly simulating the response of plastic scintillator detectors~\cite{Koh12}.  Distortions of the experimental decay energy spectra are present due to the efficiency and acceptance of the detector setup [see the inserts of Figs.~\ref{f:13Li}(b) and \ref{f:12Li}(a)].  The accuracy of the Monte Carlo simulation was estimated through varying input parameters, such as the momentum distribution of the residual $^{11}$Li fragment, and examining how this changed the efficiency curves.  The results showed that large variations of the input parameters resulted in minimal changes (less than 5$\%$) to the shape of the efficiency curves.

In addition to the three-body decay energy ($^{13}$Li), the simulations had to consistently describe other observables, such as the decay of $^{12}$Li, which could be directly populated from $pn$-stripping, and the overall multiplicity distribution of interactions in MoNA.  Limiting the data to multiplicity~$=$~1 events should enhance the contributions of the direct $^{12}$Li population. Fig.~\ref{f:12Li} shows the overall two-body ($^{11}$Li+$n$) decay energy spectrum (a), the multiplicity~=~1 gated $^{11}$Li+$n$ decay energy spectrum (b), as well as the multiplicity distribution (c) for events in coincidence with $^{11}$Li.  The two-body decay energy was calculated as $E_{\mathrm{Decay}} = M_{^{12}\mathrm{Li}} - M_{^{11}\mathrm{Li}} - M_{\mathrm{n}}$, where the invariant $^{12}$Li mass was calculated from the detected $^{11}$Li fragment and neutron.

%$E_{\rm Decay} = \sqrt{m^{2}_{f} + m^{2}_{n} + 2(E_{f}E_{n} - p_{f}p_{n}\cos(\theta_{f,n}))} - m_{f} - m_{n}$, where the subscripted $f$ and $n$ refer to the $^{11}$Li and neutron, respectively.

The Monte Carlo simulation included two components: (1) the decay of $^{13}$Li from the one-proton removal reaction and (2) the decay of $^{12}$Li from $pn$-stripping.  The black solid lines in Figs.~\ref{f:13Li} and \ref{f:12Li} represent the sum of the $^{13}$Li and $^{12}$Li contributions to the corresponding spectra.  As shown in Fig.~\ref{f:13Li}(a), a component of the decay of $^{12}$Li$\rightarrow^{11}$Li+$n$ (blue dot-dashed line) is present in the three-body decay energy spectrum from events in which the single neutron interacted twice in MoNA and thus added false two neutron events to the $^{13}$Li spectrum. Applying the causality cuts to Fig.~\ref{f:13Li}(a) should greatly reduce the presence of false two neutron events in the three-body decay spectrum. The effectiveness of the causality cuts can be seen from the near absence of the $^{12}$Li component in Fig.~\ref{f:13Li}(b).  Similarly, in the multiplicity~$=$~1 two-body decay spectrum, Fig.~\ref{f:12Li}(b), a component of the  $^{13}$Li$\rightarrow^{11}$Li+$2n$ decay (red dashed line) is present from events in which only one neutron is detected.  %Due to the relatively low efficiency for two neutron detection, the multiplicity~$=$~1 spectrum is still dominated by the $^{13}$Li contributions.

\begin{figure}
	\centering
	\includegraphics[width=0.47\textwidth]{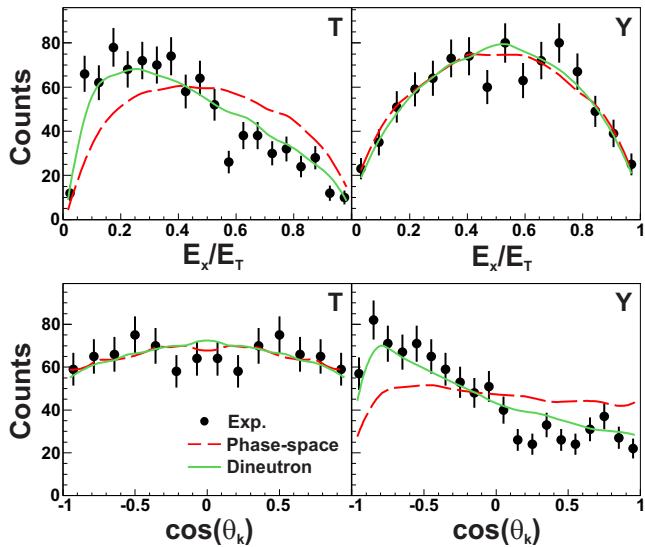}
	\caption{(color online) The relative energy ($E_{x}/E_{T}$) and angle (cos($\theta_{k}$)) as defined in the \textbf{T} and \textbf{Y} systems are shown for the experimental data (black filled circles) with causality cuts, three-body phase space decay simulation (red-dashed line), and dineutron decay simulation (green solid line).
\label{f:Jacobian}}
\end{figure}

The correlations of the three-body $^{11}$Li+$n$+$n$ system can provide additional insight into the structure and decay of $^{13}$Li.  A complete description of the three-body correlations can be obtained from the relative energy ($E_{x}/E_{T}$) and the angle ($\theta_{k}$) calculated within the \textbf{T} and \textbf{Y} Jacobi systems.  The relative energy is defined as the energy of the two-body system (frag+$n$ or $n$+$n$), $E_{x}$, relative to the total three-body energy, E$_{T}$.   A detailed description and illustration of the Jacobi coordinate systems can be found in Refs.~\cite{Char11,Ers10,Gri09,Gri01}.  In Fig.~\ref{f:Jacobian} the relative energy and angle for both the \textbf{T} and \textbf{Y} systems are shown from the experimental data with the causality cuts applied and compared to a three-body phase space and dineutron simulated emission.  The three-body phase space distribution was calculated according to Refs.~\cite{3body,3bodyROOT} with a Breit Wigner lineshape used to fit the three-body decay energy.  Further details of the dineutron decay calculation are provided later in the text.

The relative energy in the \textbf{T} system and angle in the \textbf{Y} system are particularly sensitive to the decay mechanisms, three-body phase space versus dineutron.  In the \textbf{T} system $E_{x}/E_{T}$ represents the energy of the $n$-$n$ system relative to the total three-body energy.  In the \textbf{Y} system $\theta_{k}$ is the angle between the frag+$n$ center-of-mass and the additional neutron.  The difference between the three-body phase space and dineutron decay simulations can be clearly seen in the top-left and bottom-right panels of Fig.~\ref{f:Jacobian}.  While the phase space emission (red dashed line) does not fit the experimental data, the dineutron decay (green solid line) agrees well with the data showing a strongly correlated $n-n$ emission.  This presents the second case (the first being $^{16}$Be~\cite{Spy12}) in which a ground state dineutron-like emission has been observed.  A strong dineutron component ($\sim50\%$) in the decay of $^{13}$Li for $E_{\mathrm{Decay}} > 3$~MeV was also reported by Johansson \emph{et al.}~\cite{Joh10a}.
%In both cases, this is only possible due to the lack of intermediate states providing the necessary conditions for true two-nucleon emission.

\par
Since the three-body correlations indicate that the dominate decay channel of $^{13}$Li is the emission of a dineutron (or two strongly correlated neutrons), the two-body and three-body decay energy spectra need to be analyzed accordingly.  Thus, the decay of $^{13}$Li was simulated as the emission of a dineutron, while the $^{12}$Li decay was simulated through the emission of a $s$-wave neutron with an associated scattering length.

The dineutron decay was simulated as a two-step process following the formulism of Ref.~\cite{Vol06} where the dineutron was emitted with energy $E_y$ (in the \textbf{T}, ``cluster'' system) from the $^{13}$Li followed by the decay of the dineutron with energy $E_{x}$ ($\mathrm{^{13}Li}\rightarrow \mathrm{^{11}Li} + \mathrm{^{2}n} \rightarrow \mathrm{^{11}Li}+\mathrm{n}+\mathrm{n}$).  The total amplitude for the two-step dineutron decay process is constructed from the one-body decay amplitudes as a second order process~\cite{VolyaEPJ}:
\begin{equation}
A(E_{y},E_{x})=\frac{A_{1}(E_y)\, A_{2}(E_x)}{E_{x}-\left(E_{V}-\frac{i}{2}\Gamma_{V}(E_{x})\right)},\label{eq:2b_amp_onestate}
\end{equation}
here $A_{1}(E_{y})$ and $A_{2}(E_{x})$ represent the amplitudes of the dineutron emission with kinetic energy $E_{y}$ and the subsequent dineutron breakup with intrinsic energy $E_{x}$, respectively. $E_{V}$ and $\Gamma_{V}(E_{x})$ represent the energy and width parameters of the propagator describing the dineutron virtual state. These parameters along with $A_{2}(E_{x})$ are taken from the low-energy $n-n$ scattering theory in free space, where the scattering length $a_{s}=-18.7$ fm~\cite{Gon06}, assuming that the dineutron is a $^{1}s_{0}$ $n-n$ state.

The Fermi Golden Rule gives the partial decay width distribution for the sequential process as
\begin{equation} \frac{d\Gamma(E_{T})}{dE_{y}dE_{x}}=2\pi\delta(E_{T}-E_{y}-E_{x})\left|A(E_{y},E_{x})\right|^{2}\label{eq:seq_decay}
\end{equation}
where $E_{T}$, again, represents the total energy of the three-body system.  The probability distribution then follows the usual Breit-Wigner form
\begin{equation} \frac{dP(E_{y},E_{x})}{dE_{y}dE_{x}}\propto\frac{1}{(E_{T}-E_{r})^{2}+\Gamma_{r}^{2}(E_{T})/4}\,\frac{d\Gamma(E_{T})}{dE_{y}dE_{x}},\label{eq:prob_dist}
\end{equation}
where $E_{r}$ and $\Gamma_{r}(E)$ are the resonance energy and total width of the initial state in $^{13}$Li. The delta-function in Eq.~(\ref{eq:seq_decay}) enforces that $E_{T}=E_{y}+E_{x}$ in Eq.~(\ref{eq:prob_dist}).  From the presented formalism, the $^{13}$Li decay was simulated as the emission of a dineutron with kinetic energy $E_{y}$ which proceeded to breakup with energy $E_{x}$.

The decay of $^{12}$Li, which was populated through a $pn-$removal, was simulated using a $s$-wave lineshape calculated from the model of Ref.~\cite{Bla07}.  The potential parameters used in Ref.~\cite{Bla07} to describe the $^{14}$Be and $^{14}$B projectiles are used in the present work.  This calculation was also used to describe the $s$-wave lineshape of the $^{12}$Li decay populated from the $^{14}$B(-2$p$) reaction of Ref.~\cite{Hal10}.

The four decay energy spectra of Figs.~\ref{f:13Li} and~\ref{f:12Li} were fit simultaneously varying the $^{13}$Li resonance parameters [$E_{r}$ and $\Gamma_{r}(E)$ in Eq.~(\ref{eq:prob_dist})] and the $^{12}$Li $s$-wave scattering length.  The simultaneous fitting of all the three- and two-body decay spectra greatly increased the sensitivity and robustness of the fit in comparison to individual fits of each spectrum.  The best overall fit ($\chi^{2}$ minimization) was achieved with a resonance state in $^{13}$Li at $E_{1}$~=~120$^{+60}_{-80}$~keV with $\Gamma_{r}$~=~125$^{+60}_{-40}$~keV.  A limit on the scattering length of $a_{s} > -4$ fm for the $^{12}$Li $s$-wave was determined from the $\chi^{2}$ fit which is shown in the insert of Fig.~\ref{f:12Li}(b). A reduced chi-square ($\chi^{2}/\nu$) of 1.18 was obtained from the fit with 181 degrees of freedom.  The mass excess of $^{13}$Li can be determined from the 120$^{+60}_{-80}$~keV resonance and the mass excess of $^{11}$Li (40.72828(64) MeV~\cite{Smi08}) to be 56.99(8)~MeV.

The extracted scattering length limit of $a_{s} > -4$ fm for $^{12}$Li from the present data seems to be inconsistent with the previous result of $-13.7(1.6)$~fm by Aksyutina \emph{et al.}~\cite{Aks08}.  While Aksyutina \emph{et al.} used the calculation presented in Ref.~\cite{Aks08} to calculate the $s$-wave lineshape in comparison to Ref.~\cite{Bla07} used in the present work, it was verified that the difference in the calculated lineshapes could not account for the difference in the extracted scattering lengths. However, one possible explanation for the difference would be that the events from the unidentified low-energy peak in $^{13}$Li, due to the zero two-neutron efficiency below 200 keV, were included in the $^{12}$Li spectrum of Ref.~\cite{Aks08}. This would enhance the yield at low energy thus shifting the extracted scattering length of the fit to larger negative values.

\begin{figure}
	\centering
	\includegraphics[width=0.35\textwidth]{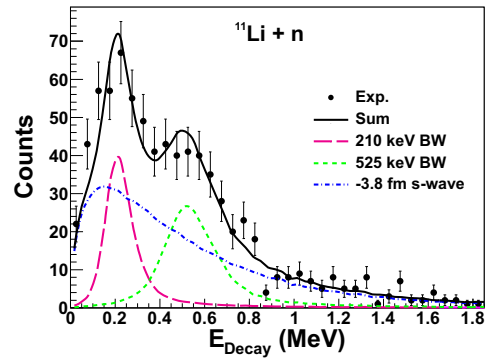}
	\caption{(color online) Measured decay energy spectrum of $^{12}$Li populated in the two-proton removal reaction from $^{14}$B.  Three components are shown to fit the data: two Breit-Wigner resonances (BW) and a $-3.8$ fm scattering length $s$-wave, corresponding to the extracted $a_{s} > -4$~fm limit from the current work.  Experimental data was taken from Ref.~\cite{Hal10}.
\label{f:Old-12Li}}
\end{figure}

A previous experiment populating $^{12}$Li~\cite{Hal10} in a two-proton removal reaction from $^{14}$B included the -13.7~fm scattering length $s$-wave from Aksyutina \emph{et al.}~\cite{Aks08} along with two narrow resonances in the fit of the $^{12}$Li decay energy spectrum.  In order to have a consistent analysis, the $^{12}$Li decay energy spectrum from Ref.~\cite{Hal10} has to be re-fit using an $s$-wave scattering length greater than $-4$ fm, as extracted in the current work.  Fig.~\ref{f:Old-12Li} shows the results of the fit ($\chi^{2}/\nu$ = 1.08) where the $s$-wave was fixed at a scattering length of $-3.8$~fm, while the resonance parameters of the other two previously observed resonances were free fit parameters. The values of the lower (pink long-dashed lines) and higher (green short-dashed line) resonances had to be adjusted slightly from 250(20)~keV to 210(30)~keV and from 555(20)~keV to 525(25), respectively.

\par
One-neutron removal experiments and shell model calculations have indicated that the $^{14}$B ground state has a dominant ($64\%-89\%$) $s_{1/2}$ configuration~\cite{Bla07,Tar04,Sav04,Gui00}.  Thus, a strong population of the $^{12}$Li $s$-wave from $^{14}$B would be expected along with a smaller population of $d$-wave resonances.  While the fit shown in Fig.~\ref{f:Old-12Li} has a 50$\%$ $s$-wave component, the statistical uncertainty of the data allows for fits of similar quality (albeit having slightly larger $\chi^{2}$) to be obtained with $s$-wave components ranging from 45$\%$ to 90$\%$, which is in agreement with $s$ and $d$ components of the $^{14}$B ground state.  In populating $^{12}$Li from $^{14}$Be(-$pn$), inclusion of the two narrow resonances in the fit of Figs.~\ref{f:13Li} and~\ref{f:12Li} did not have a significant effect and slightly increased the $\chi^{2}/\nu$ to 1.2.  Thus, the narrow resonances were not included in the presented fit of Figs.~\ref{f:13Li} and~\ref{f:12Li}. The results suggest that $^{14}$Be must have a dominant $s_{1/2}$ configuration, which is in agreement with the Coulomb dissociation measurements of Labiche \emph{et al.}~\cite{Lab01}.

\begin{figure}
	\centering
	\includegraphics[width=.32\textwidth]{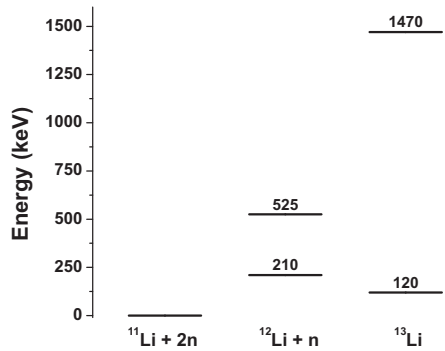}
	\caption{ Level scheme of $^{13}$Li and $^{12}$Li determined from the current work.  The 1470 keV resonance in $^{13}$Li is from Ref.~\cite{Aks08}.  The broad $s$-state with $a_{s} > -4$~fm in $^{12}$Li is not shown.
\label{f:Level}}
\end{figure}

The new level schemes of $^{13}$Li and $^{12}$Li, based on the present data and the data of Refs.~\cite{Aks08} and~\cite{Hal10}, are shown in Fig.~ \ref{f:Level}. The present experiment has very low detection efficiency for two neutron events above 1 MeV and is thus insensitive to the previously observed level at 1.47 MeV in $^{13}$Li~\cite{Aks08}.  The $^{12}$Li scattering length is not shown in Fig.~\ref{f:Level} since it corresponds to a very broad distribution in excitation energy~\cite{Bau12}. Thus the state at 210~keV has to be considered the ground state of $^{12}$Li. The mass excess of $^{12}$Li is then 49.01(3)~MeV using the most recent mass measurements of $^{11}$Li~\cite{Smi08}.

The 210~keV ground state of $^{12}$Li produces a scenario in which $^{13}$Li is bound with respect to $1n$ emission and unbound with respect to $2n$ emission.  The decay of the $^{13}$Li ground state is, thus, a genuine three-body process and, while a purely sequential decay through the $n-n$ system is unlikely, the three-body correlations indicate the presence of the $n-n$ virtual state ($a_{s}$ = -18.7~fm).  The recent work by I.~A. Egorova \emph{et al.}~\cite{Ego12} has demonstrated the complexity present in the three-body dynamics of the two-proton decay of $^{6}$Be through the use of a three-cluster model and similar calculations could provide new insights into the two-neutron emission from $^{13}$Li and $^{16}$Be~\cite{Spy12}.

%A true sequential decay through the tail of the broad $s$-state in $^{12}$Li is unlikely due to the extremely short lifetime of the state~\cite{Char11,Gri01}.
%Following Ref.~\cite{Char11}, the separation ($d_{E}$) of the neutrons at the time of the intermediate state decay can be estimated and compared to the average nuclear diameter of roughly 5 fm, which would represent a sequential process in which little influence between the neutrons occur.  For the $^{13}$Li decay through the $a_{s} > -4$ fm $s$-state, $d_{E}$ is estimated to be around 0.3 fm indicating that the emission would be nearly simultaneous.  The decay of $^{13}$Li is, thus, a genuine three-body process and, while a purely sequential decay through the $n-n$ system is unlikely, the three-body correlations indicate the presence of the $n-n$ virtual state ($a_{s}$ = -18.7~fm).

In summary, the population and decay of $^{13}$Li and $^{12}$Li were measured in $p-$ and $pn-$removal reactions from $^{14}$Be. A low-lying resonance of 120$^{+60}_{-80}$~keV was observed in $^{13}$Li for the first time, corresponding to a mass excess of 56.99(8)~MeV. Within a consistent description of all measured decay spectra the decay of $^{12}$Li could be described by an $s$-wave with a scattering length of greater than $-4$~fm, which differs significantly from the $-13.7(1.6)$~fm scattering length extracted in Ref.~\cite{Aks08}.  The $a_{s} > -4$ fm scattering length limit, from this work, was used to re-fit the decay energy spectrum of $^{12}$Li populated with a two-proton removal reaction from $^{14}$B~\cite{Hal10}.  The observed 210~keV resonance in $^{12}$Li represents the ground state and corresponds to a mass excess of 49.01(3)~MeV.  This $^{12}$Li ground state produces conditions favorable for true two-nucleon emission from $^{13}$Li to $^{11}$Li.  The angular and energy correlations in the three-body system of $^{13}$Li showed strong dineutron character for the ground state decay.

\par
The authors gratefully acknowledge the support of the NSCL operations staff for providing a high quality beam.  This material is based upon work supported by the Department of Energy National Nuclear Security Administration under Award Number DE-NA0000979 and DOE Award number DE-FG02-92ER40750.  This work was also supported by the National Science Foundation under Grant Nos. PHY06-06007, PHY09-69058, and PHY11-02511.

%\clearpage %Just because of unusual number of tables stacked at end
%\bibliography{bibdata}% Produces the bibliography via BibTeX.

% Create the reference section using BibTeX:
%\bibliography{prc_23o}

\end{document}